\documentstyle[12pt]{article}
\def\etal{{\it et al\/}}

\begin{document}
\begin{titlepage}
\begin{flushright}
  DPNU-93-32
\end{flushright}
\begin{center}
\vfill
{\large\bf Fate of Inhomogeneity in Schwarzschild-deSitter Space-time}
\vfill
{\bf Yasusada Nambu\footnote[2]{e-mail:nambu@jpnyitp}}
\vskip0.5cm
    Department of Physics, Nagoya University \\
      Chikusa, Nagoya 464-01, Japan
\end{center}
\vfill
\begin{center}
{\bf Abstract}
\end{center}
\begin{quote}
  We investigate the global structure of the space time with a
   spherically symmetric inhomogeneity using a metric junction,
 and classify all  possible
    types. We found that a motion with a negative gravitational mass
 is possible although the energy condition of the matter is not violated.
  Using the result,
     formation of black hole and worm hole during the inflationary
    era is discussed.
\end{quote}
\vfill
\end{titlepage}
%%%%%%%%%%%%%%%%%%%%%%%%%%%%%%%%%%%%%%%%%%%%%%%%%%%%%%%%%%%%%%%%%%%%%%%%%
\section{Introduction}
  The inflationary scenario is a favorable model to
  explain the homogeneity and isotropy of the present universe.
  In this scenario,  vacuum energy of
  the matter field plays a role of the cosmological constant, and the
  universe enters the phase of deSitter expansion. Initial
  inhomogeneity of the universe dumps due to the rapid cosmological
  expansion. To utilize these aspects of inflation, it is necessary to
  discuss whether the universe can enter the inflationary phase from
  the wide range of the initial condition. ``Cosmic no hair
  conjecture''  states that if a positive cosmological constant
  exists, all space-times approach deSitter space-time asymptotically.
  But it is difficult to prove and formulate  this conjecture for
  general situation and we do not know whether it is true.

 For  spherical symmetric space-time with cosmological
 constants, it is shown that the space-time does not necessarily
 approaches deSitter but a black hole and a worm hole may be
 created\cite{maeda,worm}. The global structure of the final space-time
 depends the scale and the amplitude of the initial inhomogeneity and
 this means that ``no hair conjecture'' is not necessarily established.
 But in a practical sense, a whole universe does not need to inflate
 and only a portion of our universe has to inflate(weak no hair
conjecture)\cite{gar}.
%We can say that inflation works well if the
%``measure'' of the initial condition, from which the universe has a
% inflating region, is large.

 The problem is more complicated in the early stage of the universe
because quantum effect of the matter becomes important. Even though
the completely homogeneous universe is
 created  at beginning, the inhomogeneity is continuously
 created by quantum fluctuation of the matter field in deSitter phase.
Self-reproduction of the inflating region occurs and the universe
becomes inhomogeneous on super large scale.

 In this paper, we investigate the evolution of the
inhomogeneity in the Schwarzschild-deSitter space time. As a source of
inhomogeneity, we use a false vacuum bubble with thin wall
approximation. This is the  simplest model to represent the spherically
symmetric inhomogeneity. Assuming that the inhomogeneity is created by
quantum fluctuation of the scalar field in inflationary phase, we
calculate the probability of black hole and worm hole formation. If
the probability of black hole formation is too large, the universe
becomes much inhomogeneous even though it started from a homogeneous
initial condition.

%%%%%%%%%%%%%%%%%%%%%%%%%%%%%%%%%%%%%%%%%%%%%%%%%%%%%%%%%%%%%%%%%%%%%%%%%%%%
\section{Metric Junction}
 We assume that the inside space-time of the bubble is described by
 deSitter metric with cosmological constant $ \Lambda_1$, the outside is
 described by Schwarzschild-deSitter metric with a gravitational mass $M$ and
 cosmological constant $\Lambda_2$. Using the static coordinate
 systems, they are written by
 \begin{eqnarray}
   ds^2_{in}& = & -(1-\chi^2_1 r^2)dt^2+(1-\chi^2_1 r^2)^{-1}dr^2+r^2
            d\Omega^2, \\
   ds^2_{out}&= & -(1-\frac{r_g}{r}-\chi^2_2
       r^2)dt^2+(1-\frac{r_g}{r}-\chi^2_2 r^2)^{-1} dr^2+r^2 d\Omega^2 ,
 \end{eqnarray}
 where $\chi^2_1=8\pi G\Lambda_1/3$, $\chi^2_2=8\pi G\Lambda_2/3$ and
 $r_g=2 GM$.
The motion of the bubble can be determined by the metric junction condition:
\begin{equation}
  K^i_j(in)-K^i_j(out)=4\pi\sigma G \delta^i_j,
\end{equation}
 where $K^i_j$ is the extrinsic curvature of $(2+1)$-dimensional
 hyper-surface swept out by the domain wall. $\sigma$ is the surface
 energy of the bubble and is a constant for the scalar field domain
 wall.  $\theta$-$\theta$ component
 of this equation gives the equation of motion of the bubbles radius
 $r(\tau)$ where $\tau$ is the proper time on the wall. By introducing
 the dimension-less variables\cite{blau}, our basic equation becomes
 \begin{equation}
   \left(\frac{dz}{d\tau^{'}}\right)^2+V(z)=E, \label{eqn:pot}
 \end{equation}
 where
 \begin{eqnarray}
   z&&=r/r_0, ~~~r_0^3={2 G|M|}/{\chi_{+}^2},
   ~~~\tau^{'}=\chi_{+}^2\tau/(2\kappa),
 ~~~\kappa=4\pi  G\sigma \nonumber\\
   \chi_{+}^2&&=\left[(\kappa^2+\chi^2-\chi_2^2)^2
      +(2\kappa\chi_2)^2\right]^{1/2}   ,  \nonumber\\
   V(z)&&=-\left(z-\frac{1}{z}\right)^2-{\gamma^2}/{z},\nonumber\\
   \gamma^2&&=2+2
   sgn(M)(\kappa^2+\chi_2^2-\chi^2)/{\chi_{+}^2},\nonumber\\
   E&&=-\left(\frac{2\kappa}{\chi_{+}}\right)^2(2G|M|\chi_{+})^{-2/3}.
     \nonumber
 \end{eqnarray}
 $\theta$-$\theta$ component of the extrinsic curvature of the bubble
 interior and exterior are given by
 \begin{eqnarray}
   \beta_{in}&&=\left(\frac{G|M|}{r_0^2\kappa}\right)\frac{1}{z^2}
     \left[sign(M)-\left(\frac{\chi^2
       -\chi_2^2-\kappa^2}{\chi_{+}^2}\right)z^3\right],
                          \label{eqn:betain}\\
   \beta_{out}&&=\left(\frac{G|M|}{r_0^2\kappa}\right)\frac{1}{z^2}
     \left[sign(M)-\left(\frac{\chi^2
       -\chi_2^2+\kappa^2}{\chi_{+}^2}\right)z^3\right]. \label{eqn:betaou}
 \end{eqnarray}
%%%%
The location of horizons in Schwarzschild-deSitter space is determined
by the equation
\begin{equation}
  1-\frac{2GM}{r}-\chi_2^2r^2=0,
\end{equation}
and using dimension-less variables, it can be written
\begin{equation}
  E=-sign(M)\left(\frac{4\kappa^2}{\chi_{+}^2}\right)\frac{1}{z}+
           \left(\frac{4\kappa^2\chi_2^2}{\chi_{+}^4}\right)z^2.
\end{equation}
The global structure of space-time is determined by solving
1-dimensional particle motion with the potential $V(z)$(Fig.1). The
curve of horizon is tangent to the curve  $V(z)$ at $z=z_s$. The sign
of the extrinsic curvature $\beta_{out}$ changes at this point.
The location of horizons is the intersection of $E=const.$ line and
the horizon line. For $M>0$, there are two horizons at most. As $|E|$
decreases, horizon disappears. For $M<0$, one horizon always exits.
There are three characteristic energy level that determines the
behavior of the bubble. $E1$ is the value that the horizon line
becomes maximum(at $z=z_*$) and exists for the case of positive mass.
$E2$ is the maximum value of the potential $V(z)$(at $z=z_m$). $E3$ is the
value of the potential at which the extrinsic curvature $\beta_{out}$ changes
its sign(at $z=z_s$).


\section{Classification of the space-time}
 We can completely classify the global structure of the space-time.
 The results are shown in Fig.2 and Fig.3. Fig.2 is the parameter
 space $(2GM\chi_2,\chi_1/\chi_2)$. The horizontal axis ($\chi_1/\chi_2$)
 is the scale of the inhomogeneity and the vertical axis($2GM\chi_2$)
 is the amplitude of the inhomogeneity. The parameter space is divided
 to regions R1-R14, and each region has the different global
 structure(Fig.3). As the Schwarzschild-deSitter side gives non-trivial global
  structure, we only draw this side.
 R1-R9 have the positive mass and R10-R14 have the negative mass.
  In R1 and R2, Schwarzschild horizon disappears and
  the space-time becomes deSitter like. R3 and R4 are also deSitter
  like space-time.  R5 and R9 are a worm hole space-time.
   R6,R7,R8 correspond to a black hole space-time.
{}From a viewpoint of ``weak no hair conjecture'', a worm hole space time
is ``safe'' because a bubble region does not meet a singularity in future.
For the negative mass(R10-R14), the singularity becomes time like and
 only a deSitter horizon exits. Therefore space-times are all deSitter like.
 If we take the
  limit  $\kappa\rightarrow 0$, the motion of the  bubble becomes
  null and we get the same result of Maeda \etal\cite{maeda}.  In this
  limit, R6,R7,R9,R11 and R12 disappear.

 We found that the motion with the negative gravitational mass is
 possible for the all value of $\chi_1/\chi_2$. Rewriting the junction
 condition,  the gravitational mass is given by
%%%
 \begin{equation}
  2GM=(\chi_1^2-\chi_2^2-\kappa^2)r^3+2\kappa r^2 sign(\beta_{out})
                            (1+\dot r^2-\chi_1^2 r^2)^{1/2}.   \label{eqn:mass}
 \end{equation}
%%%
The first term is the volume energy(difference between bubble interior
and exterior) and the second term is the surface energy of the bubble.
For the monotonic type solutions(R1-R5,R13,R14), we can evaluate the
above equation at
$r=0$:
\begin{equation}
  2GM=2\kappa sign(\beta_{out})r^2|\dot r|.
\end{equation}
Therefore the sign of the mass is determined by the sign of the
 extrinsic curvature of the bubble exterior. The mass can become negative
 evenif the energy condition of the matter is not violated($\kappa>0$).
 For the bounce type
 solutions(R6-R12), we
 can evaluate the mass at the turning point $\dot r=0$:
\begin{equation}
  2GM=(\chi_1^2-\chi_2^2-\kappa^2)r^3+2\kappa r^2 sign(\beta_{out})
                            (1-\chi_1^2 r^2)^{1/2}.
\end{equation}
In this case, the sign of the mass is determined by the following
characterics radius:
\begin{equation}
  r_*=\frac{2\kappa}{\sqrt{4\kappa^2\chi_1^2+(\chi_1^2-\chi_2^2-\kappa^2)^2}}.
\end{equation}
If $r>r_*$, $sign(M)=sign(\chi_1^2-\chi_2^2-\kappa^2)$(R7-R9,R10,R11)
and the sign of the mass is equal to the sign of the volume energy of
the bubble. If $r<r_*$, $sign(M)=sign(\beta_{out})$(R6,R12)

For negative mass solutions,  the singularity becomes time-like
 and is not hidden by the event horizon. But if we use the spatially
 flat time slice to foliate this space-time, this naked singularity
 does not appear to our universe.
\section{Inhomogeneity during the Inflation}
In the inflationary era, the inhomogeneity of the space time is
generated by the quantum fluctuation of the inflaton field. If the
universe is created completely homogeneous and isotropic at beginning,
later evolution is not necessarily homogeneous
because of continuous generation of the quantum fluctuation.
We discuss the probability of a black hole and a worm hole formation
during the inflation. If too many black holes are created by quantum
fluctuation, the final space time becomes much inhomogeneous and the
inflation will not succeed. We can estimate the probability of black
hole and worm hole formation within linear perturbation using the
result of the previous section(Fig.2).

The energy density is dominated by the potential energy
$V=\frac{1}{2}m^2\phi^2$ of the inflaton field and the Hubble
parameter is $\chi_0^2=8\pi/(3m_{pl}^2)V(\phi_0)$.
Let $\delta\phi_1, \delta\phi_2$ be the fluctuation of the inflaton
field interior and exterior of the bubble, respectively. $\delta\phi$
is  Gaussian random field with the average
$<\delta\phi>=0$ and the dispersion $<\delta\phi^2>=\chi^4/m^2$. The size of
the bubble is given by the horizon scale $L=\chi_0^{-1}$. The ratio of
Hubble parameters is given by
\begin{equation}
 \frac{\chi_1}{\chi_2}=\frac{\phi_0+\delta\phi_1}{\phi_0+\delta\phi_2}
     =1+\delta_1-\delta_2,
\end{equation}
where $\delta=\delta\phi/\phi$. The surface energy density is estimated to be
$\kappa=4\pi G\sigma\sim\Lambda_0/m_{pl}^2\chi_0^{-1}\sim\chi_0$, and the
velocity of the bubble is $\dot r\sim L\chi_0^{-1}=1$. The mass excess due
to the fluctuation of the inflaton field becomes
\begin{equation}
  2GM\chi_0=2(\delta_1-\delta_2)+2sign(\beta_{out})(1-\delta_1).
\end{equation}
 The probability distribution of $\delta$ is given by
\begin{equation}
  P(\delta_1,\delta_2)=N\exp\left(
      -\frac{\delta_1^2}{2<\delta_1^2>}
      -\frac{\delta_2^2}{2<\delta_2^2>}
                             \right),
\end{equation}
where $<\delta^2>=\chi_0^2/m_{pl}^2$. Combining eq.(13) and eq.(14),
the  probability of black hole
and worm hole formation for a given energy scale is obtained by monte
carlo calculation:
%%%
\begin{center}
 \begin{tabular}{|c|c|c|c|} \hline
            & BH & WH & deSitter \\ \hline
 $10^{19}GeV$ & 4.68\% & 0.15\% & 95.17\%  \\ \hline
 $10^{18}GeV$ & 0.05\%  & 0.00\% & 99.97\%  \\ \hline
 \end{tabular}
\end{center}
%%%
We can say that almost all universe becomes deSitter like.
The probability of worm hole and black hole formation increases as the
energy scale grows.
 This indicates that space-time foam structure is realized at Planck
 energy scale. Although
 the probability of black hole formation is not so small, the
characteristic mass of a black hole is small($\sim 10^{-8}$kg) at Planck
scale . It evaporate soon and does not affect later evolution of the universe.

%%%%%%%%%%%%%%%%%%%%%%%%%%%%%%%%%%%%%%%%%%%%%%%%%%%%%%%%%%%%%%%%%%%%%%%%%
\section{Summary}
We analyzed the motion of a false vacuum bubble in Schwarzschild-deSitter
space-time and obtained all possible type of motion.
The result is classified in the parameter space $(\chi_1/\chi_2,2GM\chi_2)$
which characterizes  the inhomogeneity.
Provided that the initial condition is given by random Gaussian
quantum fluctuation, we estimate the probability of black hole and worm hole
creation.

Our analysis here is limited to spherically symmetric case. But the spatial
pattern of high energy density regions by quantum fluctuation does not
necessarily have spherical shapes even though the fluctuation is
treated as perturbation.
Therefore more realistic treatment without imposing spherical symmetry
is necessary to understand correct picture of worm hole and black hole
formation via quantum fluctuation. This is our next problem\cite{nambu}.

%%%%%%%%%%%%%%%%%%%%%%%%%%%%%%%%%%
\begin{flushleft}\
\Large{\bf Acknowledgments}
\end{flushleft}
 We would like to thank S.Konno and Prof. Tomimatsu for valuable discussions.

%%%%%%%%%%%%%%%%%%%%%%%%%%%%%%%%%%%%%%%%%%%%%%%%%%%%%%%%%%%%%
%* References*
%
%%

%%%%%%%%%%%%%%%%%%%%%%%%%%%%%%%%%%%%%%%%%%%%%%%%%%%%%%%%%%%%%%%%
\begin{flushleft}
\Large{\bf Figure Captions}
\end{flushleft}
\begin{description}
\item[Fig.1]
 The shape of the potential for the wall motion. For positive mass, the wall
 intersects two times with horizon line. For negative mass, the wall crosses
 horizon only once.
\item[Fig.2]
 Classification of the type of space-time in parameter space $(\chi_1/\chi_2,
2GM\chi_2)$. The region between two dashed curves corresponds to bounce type
solution. In the limit $\kappa\rightarrow 0$, regions R6,R7,R8,R11 and R12
disappear.
\item[Fig.3]
 Trajectories of the wall in conformal diagram of space-time. deSitter space
is attached to the left side of each trajectory. Fig.3a is the case of positive
mass and Fig.3b is the case of negative mass.

\end{description}
\end{document}